\documentclass[%
 aip,
 amsmath,amssymb,
 reprint,%
]{revtex4-1}

\usepackage{graphicx}
\usepackage{dcolumn}
\usepackage{bm,bbm}

\usepackage[utf8]{inputenc}
\usepackage[T1]{fontenc}
\usepackage{mathptmx}
\usepackage{hyperref}

\newcommand{\tmu}{\tilde{\mu}}

\newcommand{\hz}{\hat{z}}

\usepackage{xcolor}

\begin{document}

\title[End exclusion zones in strongly stretched, molten polymer brushes of arbitrary shape]{End exclusion zones in strongly stretched, molten polymer brushes of arbitrary shape}

\author{Michael S. Dimitriyev}
\author{Gregory M. Grason}%
 \email{grason@mail.pse.umass.edu.}
\affiliation{ 
Department of Polymer Science and Engineering, University of Massachusetts, Amherst, Massachusetts 01003, USA
}%

\date{\today}

\begin{abstract}
Theories of strongly stretched polymer brushes, particularly the parabolic brush theory, are valuable for providing analytically tractable predictions for the thermodynamic behavior of surface-grafted polymers in a wide range of settings.  
However, the parabolic brush limit fails to describe polymers grafted to convex, curved substrates, such as the surfaces of spherical nanoparticles or the interfaces of strongly segregated block copolymers.
It has been previously shown that strongly-stretched, curved brushes require a boundary layer devoid of free chain ends, requiring modifications of the theoretical analysis.  
While this ``end exclusion zone'' has been successfully incorporated into descriptions of brushes grafted onto the outer surfaces of cylinders and spheres, the behavior of brushes on surfaces of arbitrary curvature has not yet been studied.
We present a formulation of the strong-stretching theory for molten brushes on surfaces of arbitrary curvature and identify four distinct regimes of interest for which brushes are predicted to possess end exclusion zones, notably including regimes of positive mean curvature but negative Gaussian curvature. 
Through numerical solutions of the strong-stretching brush equations, we report predicted scaling of the size of the end exclusion zone, the chain end distribution, the chain polarization, and the free energy of stretching with mean and Gaussian surface curvatures.  
Through these results, we present a comprehensive picture of how brush geometry influences the end exclusions zones and exact strong-stretching free energies, which can be applied, for example, to model the full spectrum of brush geometries encountered in block copolymer melt assembly.
\end{abstract}

\maketitle


\section{\label{sec:intro} Introduction}

\begin{figure}
\includegraphics[width=\columnwidth]{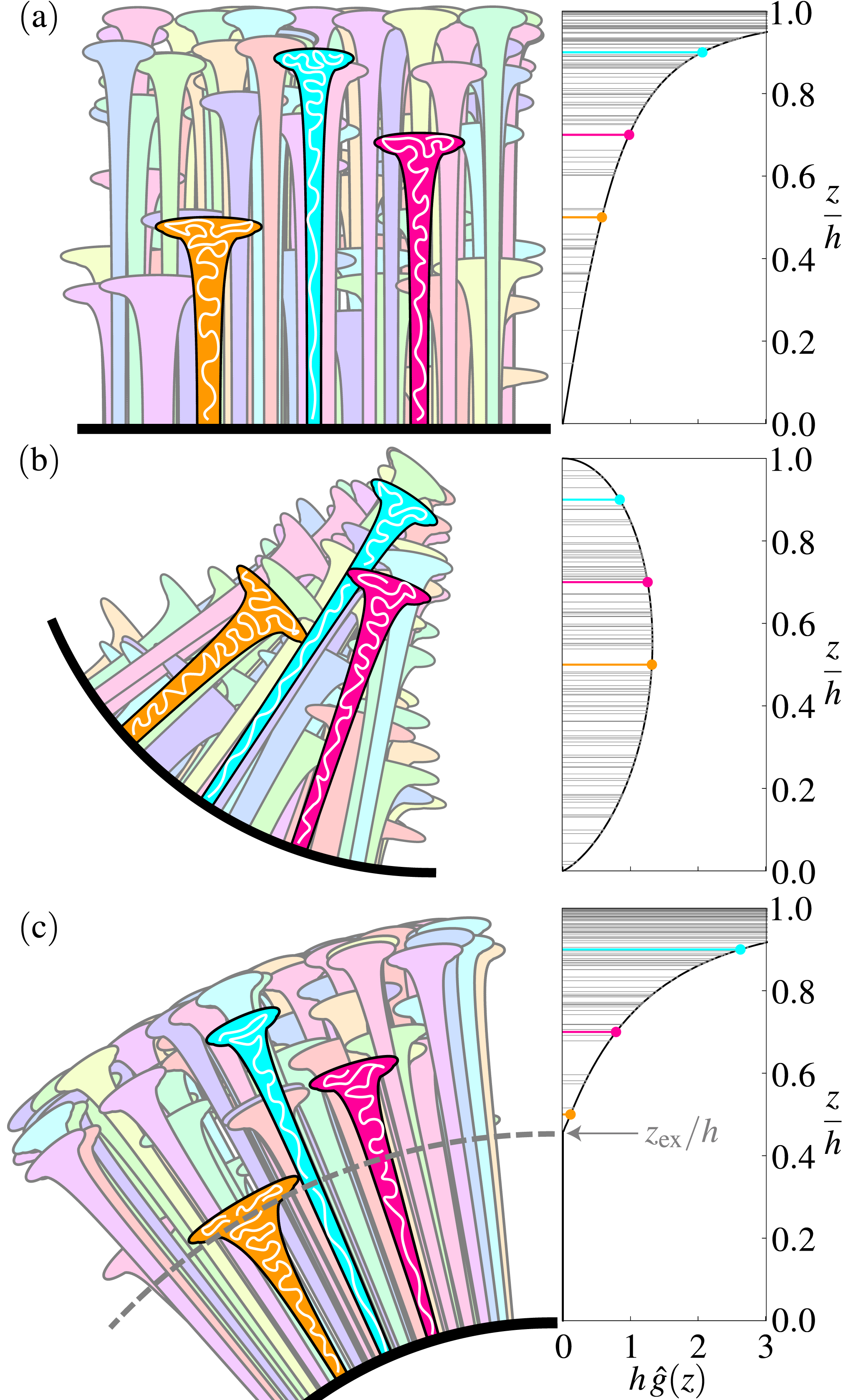}
\caption{\label{fig:1} Monomer distributions and chain heights in (a) flat, (b) converging, and (c) diverging brushes. The width of each tree-like chain profile is proportional to the number of monomers $\delta n$ occupying a thin slice $\delta z$. The distribution of chain ends is sampled from a normalized distribution function $\hat{g}(z)$ characteristic of each brush type, with three highlighted chains at height fractions $0.5 h$, $0.7 h$, and $0.9 h$. The diverging brush (c) is a cylinder of dimensionless mean curvature $Hh \simeq 1.0$ exhibits an end exclusion zone at height fraction $z_{\rm ex}/h \simeq 0.46$.}
\end{figure}

Conformations of long, flexible polymers are altered when one end is grafted to a substrate or is localized to an interface.
With increasing surface grafting density, crowding forces polymer chains to extend from the grafting surface, forming a polymer brush. 
In the strong-stretching limit of brushes, at high surface area densities, crowding increasingly distorts the volume swept out by a fluctuating polymer in the brush, eventually limiting them to regions much more narrow than their extension.  
This picture of stretching due to confinement was developed in the celebrated Alexander-de Gennes brush theory\cite{Alexander1977,deGennes1976,deGennes1980}.
While providing useful scaling laws for brush height and osmotic pressure, the Alexander-de Gennes theory is unable to capture microscopic details how chains stretch within the brush, due to unphysical constraints on the free-end positions \cite{Milner1991}.  
Relaxing the constraint that free ends are fixed at the outer edge of the brush, but considering the entropy of chains within their so-called ``classical path,'' leads to the parabolic brush theory (PBT) of strongly stretched brushes\cite{Semenov1985,Milner1988,MWC1989}.

In PBT, the mean field effects of volume interactions are captured by a spatially varying chemical potential field $\mu(z)$ that acts on segments a height $z$ above the grafting substrate.
The parabolic form of this potential, $\mu(z) - \mu(0)  \sim -z^2$, arises from the constraint that every chain has its terminal end at the substrate at $z=0$, regardless of the position of the free end.  
As a consequence, the monomer density of each chain depends on height $z$ from the substrate surface, as well as the location $z_0$ of the free end, such that chains are most strongly stretched near the substrate and are under zero tension at their free ends, so the segment distribution for each chain is strongly skewed towards the free ends.  
Schematically, this distribution of segments within chains can be illustrated as tree-like region on average with narrow ``trunks'' at anchoring surface and opening up to broad ``crowns'' at their free ends, as illustrated in Fig.~\ref{fig:1}(a).
The free-end distribution $g(z_0)$ is self-consistently coupled to the chemical potential field through volume interactions.  
For the case of molten polymer brushes considered in this paper, the form of $g(z_0)$ corresponds to the distribution of these tree-like segment profiles that integrates to a uniform total segment density. 
Since every chain in the brush connects to the substrate, every chain contributes to the monomer density of the brush near the grafting substrate, the ``floor'' of this forest of tree-like segment profiles.
The volume available to a chain near the substrate is crowded out by presence of other chains in the brush, and thus fewer chains can have free ends near the substrate.  
In contrast, fewer chains reach the upper ``canopies'' of the brush, and hence this region is more available to be filled by the free-end canopies.  
As a consequence of these arguments, in a flat (planar) brush, the probability density $g(z_0)$ of finding a chain end at height $z_0$ adopts the monotonically increasing profile displayed in Fig.~\ref{fig:1}(a).

The PBT also holds for brushes bound to concave substrates whose curvature forces chains to splay inwards, as illustrated by Fig.~\ref{fig:1}(b) for a cylindrically-curved grafting surface.
These converging brushes obey the same parabolic potential and thus adopt similar tree-shaped profiles.
However, because inward splay requires monomers to pack into regions of decreasing available area, diffuse (high density) crowns repel the tips of convergent brushes, depleting the free-end distribution $g(z_0)$, resulting in the profile shown on the right of Fig.~\ref{fig:1}(b).
It is a testament to the utility of PBT that it consistently describes both flat and converging brushes \cite{Milner1988,Grest1989,Auroy1992,Grest1994,Carignano1994,Netz1998,Manghi2001,Dimitrov2006,matsen_strong-segregation_2010}. 

However, PBT fails to consistently capture the case of chains forced to splay outwards, illustrated in Fig.~\ref{fig:1}(c).  
Because these brushes have larger fractions of their volume at their distal ends, a sufficient number of free-end crowns must be packed in this canopy.  
However, since each of these chains are rooted down at the brush floor, where there is smaller area available, the PBT leads to an overcrowding in the proximal region of the substrate, which is resolved mathematically by an unphysically negative $g(z_0)$ in this zone. 
The failure of PBT for concave brushes was recognized in early formulations by Semenov~\cite{Semenov1985}, and later resolved (first for cylindrical brushed) by generalizing the PBT solution to include an \emph{end-exclusion zone} (EEZ) within a height $z_{\rm ex}$ from the grafting substrate, where $g(0<z_0<z_{\rm ex}) = 0$~\cite{Ball1991}. 
The inclusion of an EEZ subtly alters the chemical potential $\mu(z)$ from its parabolic profile, enhancing the stretching chains near the surface, effectively redistributing more segments into the canopy zone of relatively broader crowns. 
This alteration enables a solution to the melt packing problem, resulting in the end distribution shown in Fig.~\ref{fig:1}(c).
While the EEZ extension to strong-stretching theory successfully describes brushes where PBT fails, it is not in general analytically tractable, as it solution involves a set of self-consistent integral equations coupling the chemical potential profile (and stretching) to end-zone distribution.
To date, the complexity of these equations has so far limited study to two convex geometries possible: cylinders and spheres \cite{Ball1991,Dan1992,Li1994,Belyi2004,matsen_strong-segregation_2010}.

\begin{figure}
\includegraphics[width=\columnwidth]{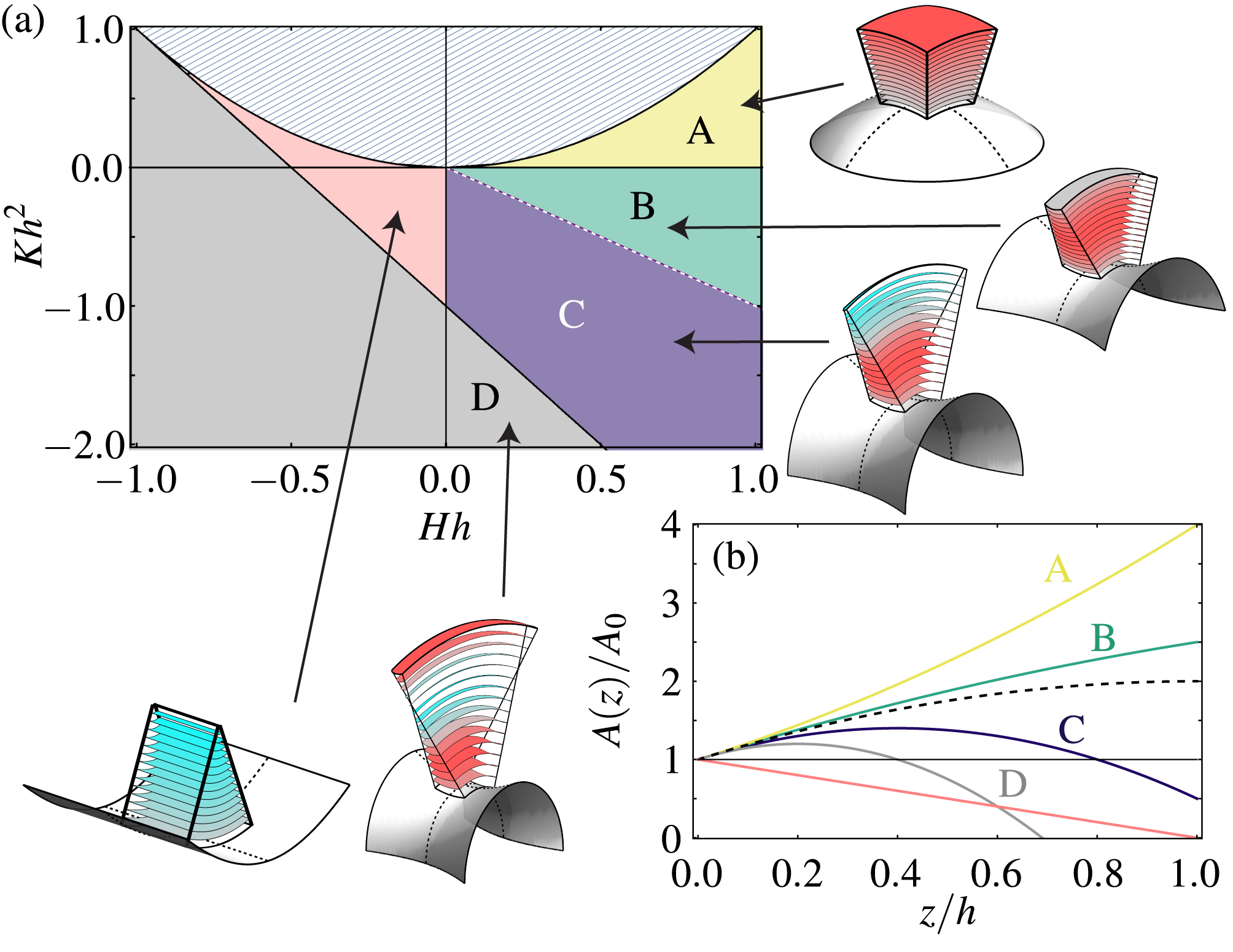}
\caption{\label{fig:2} Brushes may splay inward or outward or both and may do so anisotropically. This is controlled by the combination of substrate mean curvature $H$ and Gaussian curvature $K$, as mapped out in (a). In region A, the brush splays outward in all directions. In regions B and C, the brush splays outward in one direction and inward in the orthogonal direction. In region D, the inward splay rate overcomes the outward, causing the brush chains to pass through each other at a focal curve. Representative brush splay is illustrated by the convergence and divergence of rays extending normal to the surfaces in each of the insets. The combination of splay in different directions affects the brush area at different heights $z$ above the substrate surface, as shown in (b), and as illustrated by parallel stacks of area slices in each of the insets, with red representing area increase and blue representing area decrease relative to the area at the substrate. The boundary between regions B and C is shown as a dashed curve with the area maximum at $z^*=h$; for B, the maximum is at $z^*>h$, whereas for C, the maximum is at $z^*<h$.}
\end{figure}

In general, a grafting substrate is a 2D surface embedded in 3D and is therefore characterized by two curvatures: a mean curvature $H$ and a Gaussian curvature $K$ \cite{Hyde1997_ch1,Kamien2002}.
This combination of curvatures parameterizes the range of surfaces depicted in Fig.~\ref{fig:2}.
In the simplest case, chains are taken to emerge normal to the grafting surface, and this combination of curvatures then determines the splay of the brush.
Converging brushes emerge from negative mean curvature surfaces ($H < 0$), whereas diverging brushes emerge from positive mean curvature surfaces ($H > 0$).  
PBT consistently describes all converging brushes, since these do not require EEZs.  
At present, an extension of strong-stretching theory to incorporate EEZ has only been studied on cylinders ($K=0$) and spheres ($K=H^2$), which outline region A ($H>0$, $0\leq K \leq H^2$) in Fig.~\ref{fig:2}(a). 
Note that no surfaces exist in the regime of $K > H^2$. 
In general, surfaces in region A interpolate between cylinders and spheres, so chains always splay outward, even if the surface causes the splay to be anisotropic.
In contrast, saddle-shaped surfaces, defined by $K < 0$, force chains to splay outward in one direction and inward in the orthogonal direction.
This is the case for regions B-D and is unexplored in the context of polymer brushes.

The packing constraints in normally extended brushes are related to substrate geometry through \emph{Steiner's formula},
\begin{equation}\label{eq:steiner}
A(z) = A_0(1 + 2 H z + K z^2) \, ,
\end{equation}
where $A_0$ is the area element of the substrate and $A(z)$ is the area element at height $z$ above the substrate. 
As shown in Fig.~\ref{fig:2}(b), the area decreases for $H < 0$ and increases for $H>0$.
The increase is monotonic and accelerates in region $A$. 
However, for $K < 0$, the initial increase in area decelerates with increasing $z$, so that the area function has a maximum at $z^* = -H/K$.  
Brushes in region B have a height that is smaller than the inflection height, $h < z^*$, whereas brushes in region C are long enough that the maximum area is reached within the brush, so $h > z^*$.
For $z>z^*$, the area decreases with height, meaning that the geometry becomes convergent at its outer edge.  
Such brushes can only extend up to a maximum height where the area available to chain canopies vanishes, $A(h)=A_0(1 + 2Hh + Kh^2 )= 0$.
Region D, where $1 + 2Hh + Kh^2 < 0$, is excluded.  

A critical application of the strong-stretching of molten brushes has been the analysis of block copolymer free energies~\cite{Semenov1985, Olmsted1998, LikhtmanSemenov}.  
All domain morphologies, with the exception of lamellae, require a divergent region, which should incorporate EEZs in one of the brush-like domains.  
To date, existing strong-stretching theories of block copolymer melts are based only on PBT, and the lack of a strictly physical EEZ solutions has complicated interpretations of the these predictions.
Most notable perhaps are questions revolving around the strong-segregation stability of bicontinuous network phases, like double-gyroid, whose brush geometries include $K<0$ saddle-like shapes.  
In part, the inability to properly assess the thermodynamic stability of these so-called ``complex phases" derives the from the lack of EEZ solutions for these negatively curved regions (B and C).

In this paper, we present the solution of the self-consistent equations for strongly-stretched molten brushes in the full range of $H>0$ shapes where EEZs are required, i.e.~regions A, B and C.  
Our approach expands on the methods of Belyi \cite{Belyi2004} developed for cylindrical and spherical brushes.  
First, we present the EEZ constraint equations in a form that describes brushes of arbitrary curvature. 
We then present numerical methods for solving the EEZ constraint equations in each of the regions of interest.
Using our solutions, we find that substrate mean curvature $H$ and Gaussian curvature $K$ play distinct roles in determining how chains pack in a brush.   
These solutions show that the size and energetic costs EEZ diminishes with decreasing mean curvature of the brush (as was previously reported for cylinders and spheres), but also as Gaussian curvature becomes increasingly negative and the brush geometry evolves from divergent to convergent. 
As these results are generally applicable to systems ranging from highly curved nanoparticles beyond simple spherical and cylindrical shape \cite{Ohno2007,Dukes2010,Tung2013}, to strongly segregated block copolymer phases \cite{Milner1994,Olmsted1998,Grason2006,matsen_strong-segregation_2010}, we include a computational algorithm to interpolate our results of EEZ heights and free energies to arbitrary curvature values over the regions A, B and C in Fig.~\ref{fig:2}.

\section{\label{sec:theory} Constraint equations for curved brushes}

We consider a molten brush with a local geometry described by the area distribution of eq.~(\ref{eq:steiner}), corresponding to fixed values of curvatures $H$ and $K$ and total height $h$ (or equivalently surface density at the graft surface $\sigma_0$)~\footnote{The quadratic form of $A(z)$ need not require the {\it normal} extension of the chains in the brush.  I.e.~this form can also arise from chains trajectories that are tilted with respect to normals of the anchoring surface}.  
Each segment, indexed by $n \in [0,N]$, in a chain of $N$ total segments is located a distance $z(n)$ away from the grafting surface along the local surface normal direction.
The total free energy of a single chain in the strong-stretching is then given by the Edwards Hamiltonian \cite{Edwards1965} (in units of $k_B T$),
\begin{equation}
    \mathcal{H} = \int_0^N{\rm d}n\left\{\frac{3}{2a^2}|\partial_n z|^2 + \mu(z)\right\} \, ,
\end{equation}
where $a$ is the statistical segment length of brush chains~\cite{Matsen2002}.
The first term of $\mathcal{H}$ accounts for the entropic penalty of stretching a ``Gaussian thread'' along the $z$-direction and $\mu(z)$ is a chemical potential acting on monomers a distance $z$ from the surface.
The chemical potential $\mu$ is the mean field that enforces incompressibility (constant local density) at each point in the brush.  
Since we are considering monodisperse brushes and the density of monomers at height $z$ for a single chain is given by the ratio ${\rm d}n/{\rm d}z$, the chemical potential $\mu(z)$ must be determined in a way that is self-consistent with the statistics of single chain configurations (i.e.~the vertical distributions of free-end positions and stretching).

Following ref\cite{Milner1991}, free energy minimizing (or ``classical'') trajectories of brush chains can be understood in analogy to trajectories of a 1D particle in by an external potential $-\mu(z)$.
At time $n=0$ (the free end), the particle is released from rest, $\partial_n z|_{n=0} = 0$, at height $z_0 > 0$.
At time $n=N$, the particle must arrive at the surface at $z=0$.
The ``time-of-flight'' of such a particle can be found for an arbitrary potential $\mu$, 
\begin{equation}\label{eq:tof}
    N(z_0) = -\sqrt{\frac{3}{2a^2}}\int_{z_0}^0\frac{{\rm d}z}{\sqrt{\mu(z) - \mu(z_0)}} \, .
\end{equation}
Since every chain in the brush has the same polymerization index $N$, chain trajectories must satisfy the \emph{tautochrone constraint} $N(z_0) = N$ for all admissible values of $z_0$ \cite{AdamutiTrache1996,Hilfer2000}.
Critically, this constraint only holds for chains with free ends outside of the EEZ, i.e.~when $z_0 > z_{\rm ex}$, where the EEZ is a layer $0 \leq z \leq z_{\rm ex}$ proximal to the brush surface~\cite{Ball1991}.  
For $z_0 < z_{\rm ex}$, $N(z_0)$ is not constrained to a constant value.

Chains in a molten brush must additionally satisfy an \emph{incompressibility constraint} wherein the occupied volume fraction is constant for all $z$.
Since the brush is composed entirely of segments, the brush volume is $V = \sigma_0 A_0 N \rho^{-1}$, where $\sigma_0$ is the areal density of chains grafted to the brush substrate, $A_0$ is the area of the brush substrate, $N$ is the number of monomers per chain, and $\rho^{-1}$ is the segment volume.
This volume is constructed via parallel slices of area $A(z)$, given by eq.~(\ref{eq:steiner}) and illustrated in the insets of Fig.~\ref{fig:2}, so the number of segments in a slice of width ${\rm d}z$ is given by $\rho A(z) {\rm d}z$.
Each chain with endpoint $z_0 > z$ locally deposits ${\rm d}n(z|z_0)$ segments between $z$ and $z + {\rm d}z$. 
We can tabulate the number of chains that pass through the volume slice at $z$ by integrating endpoint distribution function $g(z_0)$ for the points $z_0\geq z$. 
Therefore, the incompressibility constraint can be expressed as
\begin{equation}\label{eq:incompressibility}
-\int_z^h{\rm d}z_0 \,g(z_0) \frac{{\rm d}n}{{\rm d}z}(z|z_0) = \rho A(z) \, ,
\end{equation}
where $g(z_0)$ is normalized such that $\int_{z_{\rm ex}}^h {\rm d}z_0\,g(z_0) = \sigma_0$.

The tautochrone eq.~(\ref{eq:tof}) and incompressibility eq.~(\ref{eq:incompressibility}) constraint equations are integral equations that are coupled by Steiner's equation eq.~(\ref{eq:steiner}).
To rewrite them in a form that is more amenable to numerical solution, following previous authors\cite{Ball1991,Dan1992,Li1994,Belyi2004}, it is convenient to change coordinates from the physical height $z$ to local value of the chemical potential $\mu$.
This change of coordinates depends on the monotonicity of $\mu(z)$, namely ${\rm d}\mu/{\rm d}z < 0$ for all $z > 0$.
At the brush surface ($z=0$), we have $\mu \equiv P$, at the EEZ boundary ($z = z_{\rm ex}$), $\mu \equiv Q$, and at the end of the brush ($z=h$), we have $\mu = 0$.
In chemical potential coordinates, eq.~(\ref{eq:tof}) is given by
\begin{equation}\label{eq:tof_mu}
N(\mu_0) = -\sqrt{\frac{3}{2 a^2}}\int_{\mu_0}^P {\rm d}\mu \frac{{\rm d}z}{{\rm d}\mu}\frac{1}{\sqrt{\mu - \mu_0}}
\end{equation}
and eq.~(\ref{eq:incompressibility}) is given by
\begin{equation}\label{eq:incompressibility_mu}
\sqrt{\frac{3}{2a^2}}\int_0^{\mu} {\rm d}\mu_0 \frac{g(\mu_0)}{\sqrt{\mu - \mu_0}} = \rho A(\mu)
\end{equation}
where $g(\mu_0) \equiv (-{\rm d}z/{\rm d}\mu)|_{\mu = \mu_0}g\left(z(\mu_0)\right)$.
This change of coordinates permits the inversion of each of these equations via an Abel transformation \cite{Ball1991,Belyi2004} so that ${\rm d}z/{\rm d\mu}$ can be solved for in terms of $N(\mu_0)$ and $g(z_0)$ can be solved for in terms of $A(\mu)$.
Through these equations, the tautochrone and incompressibility constraints in the presence of the EEZ provide non-local information about the functional forms of $z(\mu)$ and $A(\mu)$.
Here, we follow the approach of Belyi \cite{Belyi2004} and use the EEZ tautochrone constraint $N(\mu) = N$ for $\mu < Q$ to rewrite eq.~\ref{eq:tof_mu} as
\begin{equation}\label{eq:EEZ1}
\hz(\tmu < 1) = \sqrt{\frac{Q h_{\rm fl}^2}{Q_{\rm fl}h^2}}\sqrt{1 - \tmu} + \frac{1}{\pi}\int_1^{\tilde{P}} {\rm d}\tmu' \mathbbm{K}_<(\tmu,\tmu') \hz(\tmu')
\end{equation}
where $h$ is the brush height, $h_{\rm fl} \equiv \sigma_0 N \rho^{-1}$ is the height of the flat brush, $Q_{\rm fl} \equiv 3\pi^2h_{\rm fl}^2/(8 N a^2)$ is the chemical potential of the flat brush at the brush substrate, and the integral kernel $\mathbbm{K}_<$ is given by
\begin{equation}
\mathbbm{K}_<(\tmu,\tmu') \equiv \frac{\sqrt{1 - \tmu}}{(\tmu' - \tmu)\sqrt{\tmu' - 1}} \, .
\end{equation}
Here, we have introduced the scaled chemical potential coordinate $\tmu \equiv \mu/Q$ and the dimensionless height coordinate $\hz \equiv z/h$. 
Similarly, we can use EEZ incompressibility constraint $g(\mu) = 0$ for $P > \mu > Q$ to rewrite eq.~(\ref{eq:incompressibility_mu}) as
\begin{equation}\label{eq:EEZ2}
\mathcal{A}(\tmu > 1) = \frac{1}{\pi}\int_0^{1}{\rm d}\tmu'\,\mathbbm{K}_>(\tmu,\tmu')\mathcal{A}(\tmu')
\end{equation}
where the $\mathcal{A}$ is the dimensionless area $\mathcal{A}(\mu) \equiv A(\mu)/A_0$ and the integral kernel $\mathbbm{K}_>$ is given by
\begin{equation}
\mathbbm{K}_>(\tmu,\tmu') \equiv \frac{\sqrt{\tilde{\mu} - 1}}{(\tmu - \tmu')\sqrt{1-\tmu'}} \, .
\end{equation}
Notably, these two integral equations relate information about the unknown functions outside of the EEZ, $\tmu < 1$, to information about the unknown functions inside of the EEZ, $1 \leq \tmu \leq \tilde{P}$~\cite{Belyi2004}.

The EEZ constraint equations, eqs.~(\ref{eq:EEZ1}) \& (\ref{eq:EEZ2}), are related by Steiner's formula $\mathcal{A}(\tmu) = 1 + 2\hat{H}\hz(\tmu) + \hat{K}\hz^2(\tmu)$, where $\hat{H} \equiv H h$ and $\hat{K} = K h^2$ are the dimensionless mean and Gaussian curvatures, respectively.
We can, in principle, solve these coupled integral equations numerically, despite the nonlinear form of their coupling.
This has been done in the specialized cases where the curved surface is described by a single radius of curvature $R$, namely for cylinders ($H = 1/(2R)$ and $K = 0$) as well as spheres ($H = 1/R$ and $K = 1/R^2$) \cite{Belyi2004}.

To solve the equations for arbitrary curvature, we express $\hz(\tmu)$ in terms of a new function $\psi(\tmu)$ via
\begin{equation}
\hz(\mu) = \frac{1}{\alpha\hat{H}}\left(\frac{\psi(\tmu)}{\Psi} - 1\right)\, ,
\end{equation}
where 
\begin{equation}
\alpha \equiv K/H^2
\end{equation}
is a dimensionless parameter characterizing the scale of the Gaussian curvature $K$ relative to the mean curvature $H$, $\Psi$ is a dimensionless parameter that, through the boundary condition $\hat{z}(\tilde{P}) = 0$, takes on the boundary value $\Psi = \psi(\tilde{P})$.
Through Steiner's formula, we can find a relationship between $\psi(\tmu)$ and $\mathcal{A}(\tmu)$, namely
\begin{equation}\label{eq:psi_area}
\mathcal{A}(\tmu) = 1 + \frac{1}{\alpha}\left(\frac{\psi^2(\tmu)}{\Psi^2} - 1\right) \, .
\end{equation}
After substitution of $\psi(\tmu)$ into Eqs.~(\ref{eq:EEZ1}) \& (\ref{eq:EEZ2}), we are left with a pair of explicitly coupled nonlinear integral equations,
\begin{subequations}
\begin{align} \label{eq:psi_int_equations}
\psi(\tmu < 1) &= \alpha\sqrt{1-\tmu} + \frac{2}{\pi}\Psi\arctan\sqrt{\frac{1-\tmu}{\tilde{P}-1}} \notag \\
&\mkern+32mu + \frac{1}{\pi}\int_1^{\tilde{P}} {\rm d}\tmu' \mathbbm{K}_<(\tmu,\tmu')\psi(\tmu')\\
\label{eq:psi_int_equations2}
\psi^2(\tmu > 1) &= \frac{2}{\pi}\Psi^2(1 - \alpha)\arctan\sqrt{\tmu - 1} \notag\\
&\mkern+32mu + \frac{1}{\pi}\int_0^1 {\rm d}\tmu' \mathbbm{K}_>(\tmu,\tmu')\psi^2(\tmu')
\end{align}
\end{subequations}
where we have chosen $\Psi = \hat{H}^{-1}\sqrt{Q_{\rm fl}h^2/(Q h_{\rm fl}^2)}$ to remove any dependence of the integral equations on $Q$, $h$, or $\hat{H}$, leaving only $\alpha$ and $\tilde{P}$ as parameters.

The integral equations eqs.~(\ref{eq:psi_int_equations}) \& (\ref{eq:psi_int_equations2}) have quadratic coupling, similar to the integral equations describing spherical brushes\cite{Belyi2004}, yet are valid for arbitrary Gaussian curvature $K \neq 0$ (the cylindrical limit of the equations are included in the Appendix \ref{app:cyl} for completion).
Region A ($0 < \alpha \leq 1$), are described by solutions for which $\psi(\tmu)/\Psi \geq 1$ for all values of $\tmu$, with equality at $\tmu = \tilde{P}$; negative Gaussian curvature regions B-D ($\alpha < 0$) are described by solutions with $\psi(\tmu)/\Psi \leq 1$.


\section{\label{sec:numerical} Numerical Methods}

The forms of the coupled integral equations eqs.~(\ref{eq:psi_int_equations}) \& (\ref{eq:psi_int_equations2}) are reminiscent of coupled inhomogeneous Fredholm integral equations of the second kind, but their nonlinearity puts them outside of this classification, complicating the process of finding solutions.
Nevertheless, we take a classical approach of approximating the integral equations as algebraic equations via the Nystr\"om method \cite{NumericalRecipes} which we then solve iteratively.
We find that this approach works well for $(\alpha,\tilde{P})$ in region A, consistent with previous results focusing on cylinders as spheres~\cite{Belyi2004}, but also for broad parts of region C and D.
For region B, we are unable to find convergence to physical solutions, hinting at some sort of instability in the iterative procedure.
However, we are able to find approximate solutions through a variational method that we outline below.
We have included code that implements the numerical schemes discussed here in a publicly accessible repository\cite{dataset}.

\subsection{Iterative method}

For the iterative method, to ensure continuity at the EEZ, we rewrite the function $\psi$ as $\psi(\tmu) = \psi_1 + \Delta \psi(\tmu)$ and $\psi^2(\tmu) = \psi_1^2 + \Delta\psi^2(\tmu)$, where $\psi_1 \equiv \psi(\tmu = 1)$.
This allows us to rewrite eqs.~(\ref{eq:psi_int_equations}) \& (\ref{eq:psi_int_equations2}) as
\begin{subequations}
\begin{align} \label{eq:psi_int_equations_cont}
\psi(\tmu < 1) &= \psi_1 + \alpha\sqrt{1-\tmu} + \frac{2}{\pi}(\Psi-\psi_1)\arctan\sqrt{\frac{1-\tmu}{\tilde{P}-1}} \notag \\
&\mkern+32mu + \frac{1}{\pi}\int_1^{\tilde{P}} {\rm d}\tmu' \mathbbm{K}_<(\tmu,\tmu')\Delta\psi(\tmu')\\
\label{eq:psi_int_equations_cont2}
\psi^2(\tmu > 1) &= \frac{2}{\pi}\psi_1^2\arcsin\frac{1}{\sqrt{\tmu}} + \frac{2}{\pi}\Psi^2(1 - \alpha)\arctan\sqrt{\tmu - 1} \notag\\
&\mkern+32mu + \frac{1}{\pi}\int_0^1 {\rm d}\tmu' \mathbbm{K}_>(\tmu,\tmu')\Delta\psi^2(\tmu')
\end{align}
\end{subequations}
which are explicitly continuous at the EEZ boundary $\tmu = 1$.
Next, we discretize the interval $[0,\tilde{P}]$ into $\mathcal{N} = 10^4$ subdivisions so that the evaluation of $\psi(\tmu)$ results in a vector $\psi_n$ for $n = 0,1,\dots,\mathcal{N}$.
This allows us to recast eqs.~(\ref{eq:psi_int_equations_cont}) \& (\ref{eq:psi_int_equations_cont2}) as coupled algebraic equations of the form $\psi(\tmu < 1) = f_<[\psi](\tmu)$ and $\psi^2(\tmu > 1) = f_>[\psi^2](\tmu)$, where the integrals are approximated using Simpson's Rule.
The goal is then to seek zeros of the residuals $g_<[\psi](\tmu < 1) \equiv f_<[\psi](\tmu) - \psi(\tmu)$ and $g_>[\psi^2](\tmu > 1) \equiv f_>[\psi^2](\tmu) - \psi^2(\tmu)$.
We proceed to iterate these equations using simple mixing with the rules 
\begin{equation}\begin{split}
\psi^{(i+1)}(\tmu < 1) &= \psi^{(i)}(\tmu) + r g_<\left[\psi^{(i)}\right](\tmu) \\
\psi^{2,(i+1)}(\tmu > 1) &= \psi^{2,(i)}(\tmu) + r g_>\left[\psi^{2,(i)}\right](\tmu)
\end{split}\, ,\end{equation}
where the mixing parameter $r$ takes on values in the interval $(0,1]$.
We iterate until the $L^2$ norm of the total residual, defined as 
\begin{align}\label{eq:total_residual}
{\rm Res}\left[\psi_<^{(i)},\psi_>^{(i)}\right] &\equiv \int_0^1 {\rm d}\tmu g^2_<\left[\psi^{(i)}\right](\tmu)  \notag \\
&\mkern+32mu + \int_1^{\tilde{P}} {\rm d}\tmu g^2_>\left[\psi^{(i)}\right](\tmu) \, ,
\end{align}
reaches a target value.
In order to iterate these equations, the square root of eq.~(\ref{eq:psi_int_equations_cont2}) must be taken, leading to a sign ambiguity.
Per eq.~(\ref{eq:psi_area}), the maximum of the area function $\mathcal{A}$ occurs at $\psi = 0$, which occurs at a point $\tmu^* = \tmu(\hat{z}^*)$ in chemical potential coordinates.
Since the brush can pass through the maximum area at most once, the sign of $\psi(\tmu)$ is determined by the sign of $\tmu - \tmu^*$.
We use linear interpolation on $\psi^2$ to determine the location $\tmu^*$ of any zeros and then assign the sign of the square root $\psi = \pm \sqrt{\psi^2}$ accordingly.

This method works well for $\alpha > 0$, consistently reaching total residuals of $10^{-10}$ or lower.
To survey solutions in the region A, we performed sweeps in $\tilde{P}$ for $\alpha = 0.1,\, 0.2, \dots 1.0$; the results of these sweeps are discussed in the next section.
Use of the iterative method on the above equations for $\alpha < 0$ (particularly, in region B where the magnitude of $|-K|$ is small) proves to be unstable, leading to rapidly diverging residuals.
This instability seems to be influenced by the negative values of $\alpha\sqrt{1-\tmu}$ in eq.~(\ref{eq:psi_int_equations}).
It is possible to find stable solutions by instead choosing $\Psi \mapsto \Psi/\alpha$, resulting in a modified equation,
\begin{align}
\psi(\tmu < 1) &= \sqrt{1-\tmu} + \frac{2}{\pi}\Psi\arctan\sqrt{\frac{1-\tmu}{\tilde{P}-1}} \notag \\
&\mkern+32mu + \frac{1}{\pi}\int_1^{\tilde{P}} {\rm d}\tmu' \mathbbm{K}_<(\tmu,\tmu')\psi(\tmu') \, .
\end{align}
However, using values of $\alpha$ between -0.1 and -10, this only results in stably converging solutions in regions C and D and the iteration method fails to find solutions in region B.
Interestingly, while the $\tilde{P} \rightarrow 1$ limit corresponds to vanishing mean curvature $\hat{H}$ for region A, consistent with the disappearance of the EEZ, the same limit corresponds to an increasingly large magnitude of Gaussian curvature $|\hat{K}|$ for regions C and D.
This suggests that while the size of the EEZ is largely dependent on a positive mean curvature $H>0$, it is also modulated by Gaussian curvature to the degree that a sufficiently negative value $K<0$ at fixed $H$ can suppress the EEZ. 
We will show that this is indeed the case.

\subsection{Variational method}

In order to study solutions to eqs.~(\ref{eq:psi_int_equations}) \& (\ref{eq:psi_int_equations2}) in region B, we turn to a variational calculation, the goal of which is to determine forms of $\psi(\tmu)$ that minimize the total residual, eq.~(\ref{eq:total_residual}).
We approximate $\psi(\tmu)$ by a pair of degree-$d$ Bernstein polynomials,
\begin{equation}\begin{split}
\psi(\tmu < 1) &= \sum_{n=0}^d a_n \beta_n^d(\tmu) \\
\psi(\tmu > 1) &= \sum_{n=0}^d b_n \beta_n^d\left(\frac{\tmu-1}{\tilde{P}-1}\right)
\end{split}\, ,\end{equation}
where the Bernstein polynomials are defined as
\begin{equation}
\beta_n^d(t) = \frac{d!}{n!(d-n)!}(1-t)^{d-n}t^n \, ,
\end{equation}
and $a_n,b_n \in \mathbbm{R}$ are coefficients representing the degrees of freedom for the variational calculation.
We further restrict the space of functions by requiring continuity of $\psi(\tmu)$, so $b_0 = a_d$, and continuity of ${\rm d}\psi/{\rm d}\mu$, so $b_1 - b_0 = (\tilde{P}-1)(a_d - a_{d-1})$.
As a result, the variational calculation has $2d-2$ degrees of freedom.
This Bernstein basis is not only convenient due to its smoothness, it also guarantees that the function $\psi(\tmu)$ is well-behaved, despite the weakly singular kernels $\mathbbm{K}_<$ and $\mathbbm{K}_>$ \cite{Jafarian2014}.
We numerically minimize the total residual with respect to these coefficients via a Nelder-Mead method, as implemented in the \texttt{Optim} package in the Julia programming language.
While this method generally leads to higher residuals than the iteration method, we do find reasonable solutions achieving residuals of $10^{-6}$ or smaller for $d = 16$ with a gradient tolerance of $10^{-12}$.  However, we note that the performance of even this variation method is limited, such that achieving residuals of even this weaker tolerance becomes problemation for sufficiently large alpha $|\alpha|$ (i.e. near to the the $K =0$ line).
We performed sweeps of $\tilde{P}$ for $\alpha$ between -0.1 and -3.

\section{\label{sec:discussion} Results \& Discussion}

\begin{figure*}
\includegraphics[width=\textwidth]{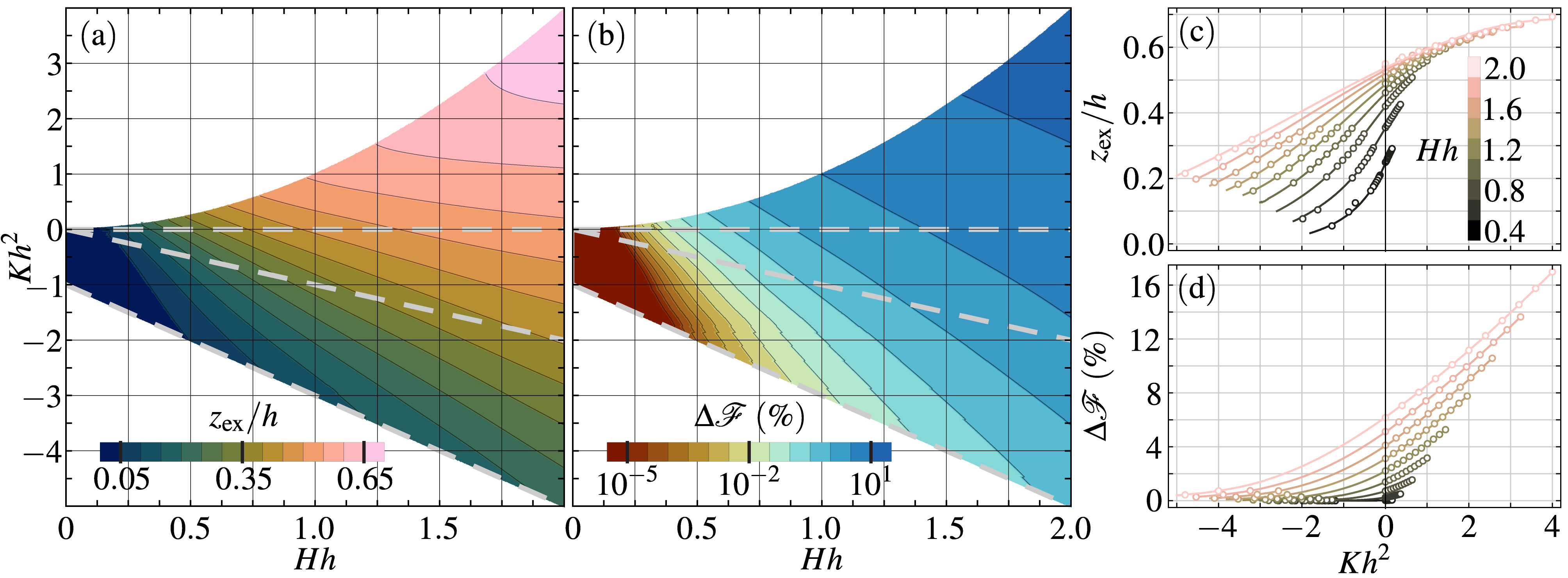}
\caption{\label{fig:3} 
Summary of numerical studies after smoothing and interpolation, plotted as a function of dimensionless mean curvature $Hh$ and Gaussian curvature $Kh^2$. We extrapolate to the low curvature regime $Hh\rightarrow 0$ using a fitting function of the form $a\times{\rm exp}(-b/H)$ for contours of fixed $Kh^2$, where $a$ and $b$ are fitting parameters. Data shown for physical regions A, B, and C, the boundaries of which are denoted with dashed lines. (a) Fraction of exclusion zone height $z_{\rm ex}$ to brush height $h$. (b) Percent correction $\Delta \mathcal{F}$ to the parabolic brush free energy due to the inclusion of the end exclusion zone, plotted on log scale. (c) and (d) show relative EEZ size and percent free energy correction as a function of Gaussian curvature $Kh^2$ for fixed values of mean curvature $Hh$ ranging from $0.4$ to $2.0$. 
}
\end{figure*}

A solution for $\psi(\tmu)$ for fixed values of $(\alpha,\tilde{P})$ fully encodes both the structure (stretching and free-end profiles) as well as thermodynamics of strongly-stretched, convex brushes.
The dimensionless mean curvature $\hat{H}$ is given by $\hat{H} = \alpha^{-1}(\psi(0)/\Psi - 1)$, from which we can find the dimensionless Gaussian curvature $\hat{K} = \alpha \hat{H}^2$.
The brush height $h$ is found through volume conservation: the volume of the flat brush, $V_{\rm fl} = A_0h_{\rm fl}$, must be the same as the volume of the curved brush, $V = \int_0^h {\rm d}z\, A(z) = A_0h(1 + \hat{H} + \hat{K}/3)$, so 
\begin{equation}
h = h_{\rm fl}(1 + \hat{H} + \hat{K}/3)^{-1} .
\end{equation}
From this, we extract the height of the EEZ, $z_{\rm ex}/h = (\psi(1) - \Psi)/(\psi(0) - \Psi)$, as well as its chemical potential, $Q/Q_{\rm fl} = h^2/(\hat{H}\Psi h_{\rm fl})^2$.
We can calculate $g(z_0)$ by first inverting eq.~(\ref{eq:incompressibility}) via an Abel transform, yielding
\begin{equation}
g(\mu_0) = \sqrt{\frac{2 a^2}{3 \pi^2}}\frac{{\rm d}}{{\rm d}\mu_0}\int_0^{\mu_0} {\rm d}\mu \frac{\rho A(\mu)}{\sqrt{\mu_0 - \mu}}
\end{equation}
and then performing a change of variables, $g(z_0) = -({\rm d}\mu/{\rm d}z)_{z = z_0}g(\mu(z_0))$.
Finally, we find it convenient to define a scaled distribution function $\hat{g}(z_0) \equiv g(z_0)/\sigma_0$, normalized such that $\int_0^h {\rm d}z_0\,\hat{g}(z_0) = 1$.

In Fig.~\ref{fig:3}(a), we plot the end exclusion zone height $z_{\rm ex}/h$ as a function of dimensionless mean curvature $Hh$ and Gaussian curvature $Kh^2$ by interpolating the results of our numerical studies.
We find that the size of the EEZ grows monotonically with $Hh$ at fixed $Kh^2$.
However, this growth rate is strongly modulated by $Kh^2$.
Recall from Fig.~\ref{fig:2}(b) and Steiner's equation eq.~(\ref{eq:steiner}) that the area $A(z)$ available to brush monomers is controlled by mean curvature $H$ for small $z$ and by Gaussian curvature $K$ for large $z$.
The strong dependence of the EEZ size on $Hh$ for low-to-intermediate curvatures can be rationalized by the proximity of the EEZ boundary layer to the grafting substrate: $Hh$ sets the packing geometry of monomers \emph{local} to the EEZ.
However, the strong modulation by $Kh^2$ indicates that the packing of monomers at the brush end exerts \emph{non-local} control over the size of the EEZ.
At high curvatures, the Gaussian curvature plays the dominant role in determining the EEZ size, with the variations in $z_{\rm ex}/h$ with $Hh$ largely saturating in the $Hh \simeq 2$ regime, as seen in Fig.~\ref{fig:3}(c).
Here, the EEZ is large enough that it starts to ``feel'' the packing geometry at the brush end more acutely.

The numerical schemes outlined here do not work well in the limit of a nearly flat brush, where $Hh\rightarrow 0$ and $\tilde{P}\rightarrow 1$.
However, as shown in Appendix \ref{app:weak}, our equations are amenable to the same low-curvature analysis as presented by Belyi \cite{Belyi2004} and we recover the same scaling prediction $z_{\rm ex}/h \sim \exp\{-(2Hh_{\rm fl})^{-1}\} \sim \exp\{-[(2Hh)(1+Hh+Kh^2/3)]^{-1}\}$.
Not only does this scaling function fit our lowest-curvature data well and serve to extrapolate our data to $Hh\rightarrow 0_+$, the variations with $Kh^2$ are also consistent with the trends observed in our data.
Interestingly, this scaling function is non-analytic in $Hh$ and thus does not permit a power series expansion about $Hh\rightarrow 0_+$.

As shown in Appendix \ref{app:fe}, the stretching free energy of a brush is given by 
\begin{equation}\label{eq:fe}
F = \frac{\rho}{2}\int_0^P {\rm d}\mu\, z(\mu) A(\mu) = \frac{\rho}{2}\int_0^h {\rm d}z \Big(- z\frac{ {\rm d} \mu}{ {\rm d} z} \Big)  A(z)  \, .
\end{equation}
Note that PBT brush result $\big(-z \frac{ {\rm d} \mu}{ {\rm d} z} \big) \propto z^2$ shows that the entropic free energy of occupying a height $z$ in the molten brush increases with that height~\cite{LikhtmanSemenov}.
The existence of the EEZ modifies the form of $\big(-\frac{ {\rm d} \mu}{ {\rm d} z} \big)$ and hence the relative free energy costs of occupying different heights from the anchoring surface are re-weighted with respect to the PBT. 
To examine the effect of the EEZ on the free energy, we determine the free energy change relative to the PBT free energy, $\Delta\mathcal{F} \equiv (F-F_{\rm PBT})/F_{\rm PBT}$, which is plotted in Fig.~\ref{fig:3}(b).
While the contours of equal free energy correction follow a qualitatively similar trend to the EEZ height in Fig.~\ref{fig:3}(a), the difference in scale is dramatic, with free energy corrections $\lesssim 10^{-4}\%$ for an EEZ height of $\simeq 10\%$ of the brush height.
Gaussian curvature similarly modulates the size of the free energy correction, which decays for negative $K$, as shown in Fig.~\ref{fig:3}(d).
A low curvature approximation of $\Delta \mathcal{F}$ in Appendix \ref{app:weak} suggests reveals a similar scaling with powers of ${\rm exp}\{-(2Hh_{\rm fl})^{-1}\}$, indicating that for low curvature, $\Delta \mathcal{F} \sim (z_{\rm ex}/h)^{2}$ with $\nu = 2$. 
However, our numerical results are unable to resolve sufficiently low curvatures to demonstrate this scaling prediction.
As the stretching free energy corrections are remarkably in the low curvature, with values of $\Delta F \gtrsim 1\%$ only appearing for $Hh\gtrsim 0.5$, low-curvature EEZ corrections are unlikely have a significant effect of brush thermodynamics.
Nevertheless, for relatively high curvatures, the EEZ correction to the free energy can be substantial.
To make these results accessible for future studies, we have included supporting software for interpolating EEZ results for arbitrary values of $Hh$ and $Kh^2$ in this range\cite{dataset}.

\begin{figure*}
\includegraphics[width=\textwidth]{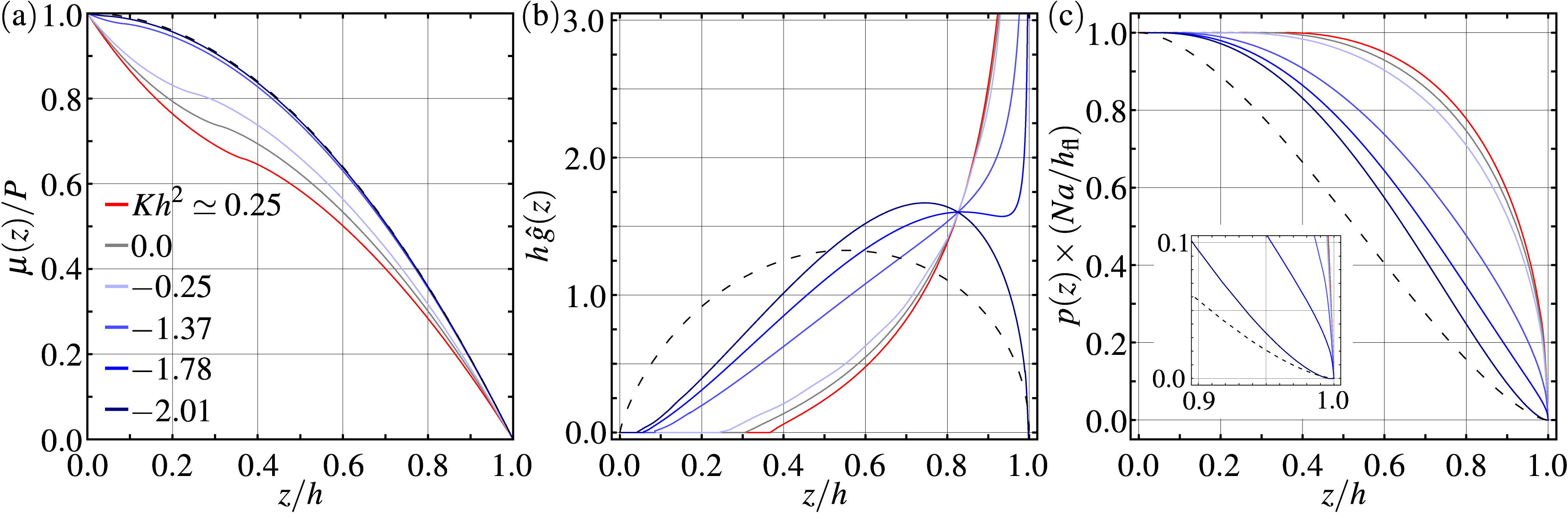}
\caption{\label{fig:4}Properties of brushes at fixed $Hh \simeq 0.5$, with $Kh^2\simeq 0.25$ representing a spherical brush, $Kh^2=0$ a cylindrical brush, $Kh^2 \simeq -0.25$ a region B brush, $Kh^2 \simeq -1.37$ and $-1.78$ two region C brushes, and $Kh^2 \simeq -2.01$ a brush near the C/D border. (a) Chemical potential function $\mu(z)$, normalized by the chemical potential at the grafting surface $\mu(0) = P$ for a subset of solutions for clarity (note: $Kh^2 \simeq -1.8$ not shown in this panel since it is indistinguishable from $Kh^2 \simeq -2$). The dashed curve corresponds to the PBT prediction for a concave cylindrical brush with $Hh=-1/2$. The end distribution function $\hat{g}(z)$ and polarization order parameter $p(z)$ for the same subset of solutions are plotted in (b) and (c). The inset of (c) is a magnified view of the polar order parameter as $z\to h$, showcasing the similarities in local chain packing between the concave cylindrical brush and the $Kh^2 \simeq -2$ brush.}
\end{figure*}

To put the magnitudes of free energy corrections in perspective, we can consider domains of AB diblock copolymer melts in their typical range of stability, which is determined by the volume fraction $f_A$ of block A relative to block B, as well as the segregation strength $\chi N$.
At finite $\chi N$, the equilibrium configurations of block copolymers are well-described by a variant of the self-consistent field theory (SCFT) \cite{Edwards1965,Helfand1975}.
Semenov\cite{Semenov1985} postulated that in the strong segregation limit ($\chi N \to \infty$), chains are strongly stretched, with each domain of the microphase separated block copoloymer melt resembling a molten brush.
To establish that this strong-segregation theory (SST) is a rigorous asymptotic limit of SCFT, Likhtman and Semenov\cite{LikhtmanSemenov} determined finite-$\chi N$ corrections based on the translational entropy of chain ends and organization of chains at the AB-interface.
Matsen\cite{matsen_strong-segregation_2010} demonstrated that EEZ effects emerge in the high-$\chi N$ regime of SCFT calculations and that an accurate comparison with SST requires incorporating EEZ corrections to the stretching free energy.
This is particularly true when establishing the stability windows for cylinder ($0.11 \lesssim f_A \lesssim 0.30$) and sphere ($f_A \lesssim 0.11$) phases in the strong-segregation limit \cite{Fredrickson2006,matsen_strong-segregation_2010}.
The corresponding curvature ranges for these two phases are $1.0 \gtrsim Hh \gtrsim 0.4$ and $Hh \gtrsim 1.1$ \& $Kh^2 \gtrsim 1.2$, respectively.
Our results show an EEZ correction to the free energy of $1.37\% \gtrsim \Delta\mathcal{F}_{\rm cyl} \gtrsim 0.03\%$ and $\Delta\mathcal{F}_{\rm sph} \gtrsim 4.17\%$.
These free energy corrections are close to the values of $1.43\% \gtrsim \Delta\mathcal{F}_{\rm cyl} \gtrsim 0.04\%$ and $\Delta\mathcal{F}_{\rm sph} \gtrsim 4.04\%$ obtained by Matsen\cite{matsen_strong-segregation_2010} in a rather different approach to determining the EEZ corrections to PBT that allows free chain ends to have variable tension.

Network phases, such as the double-gyroid, have morphologies with local surface curvatures and domain thicknesses that vary with position over the intermaterial dividing surface between minority and matrix domains.  
Taking estimates from ref.\cite{Reddy2021} the brush geometry for a double-gyroid varies from a locally flatter region near to points of 3-fold symmetry (along $\left<110\right>$ directions) with $H h \simeq 0.1 $ and $K h^2 \simeq 0$ to the most negatively curved points, at ``elbow’’-like regions that pass through points of 2-fold symmetry (along $\left<110\right>$ directions) with  $H h \simeq 0.2 $ and  $K h^2 \simeq -0.3$ (a region B brush). 
Each of these regions have free energy corrections of at most $\lesssim 10^{-6}\%$, suggesting that EEZ-based corrections are minute for such double gyroid morphology, relative to cylindrical and sphere morphologies. 
Nevertheless, given small free energy differences with competing phases, the extent to which the free energy corrections to the PBT alter the predicted phase boundaries in the strong-segregation ($\chi N \to \infty$) limit remains an open question.

To study how chains behave in each of the regions of interest, we consider solutions with $Hh\simeq 0.5$ and variable $Kh^2$ in Fig.~\ref{fig:4}, in particular how the extent of the EEZ changes.
As shown in Fig.~\ref{fig:4}(a) the chemical potential $\mu(z)$ approaches the PBT chemical potential, under suitable re-scaling by $h$ and $P$, as $Kh^2$ becomes increasingly negative, consistent with the observation that the brush free energy is well approximated by the PBT free energy in this regime.
Varying $Kh^2$ can be accomplished by either fixing $K$ and varying $h$ or vice-versa.
If $K$ is fixed to a negative value, and the brush height $h$ is increased, the size of the EEZ $z_{\rm ex}$ remains fixed but the fraction $z_{\rm ex}/h$ decreases, so that on the scale of the brush, the overall effect of the EEZ diminishes.
Conversely, if $h$ is fixed and $K$ becomes increasingly negative, the area growth due to positive $H$ is quickly overcome by the decrease in area due to negative $K$, decreasing the size $z_{\rm ex}$ of the EEZ.
Physically, as the end of the brush is confined to a smaller area, chain ends are depleted away from the end towards the middle of the brush, as shown in Fig.~\ref{fig:4}(b).
The result is a balance between packing constraints proximal near the grafting substrate and near the free brush end; as the brush becomes more confined, the outward splay near the substrate is reduced.
Thus, the non-local nature of the integral equations describing curved brushes is reflected in the non-local influence on chain packing that the two ends of the brush exert on each other.
The forms of the chemical potential $\mu(z)$ and end distribution function $\hat{g}(z)$ obtained from our numerical studies of the brush equations are strikingly similar to the SCFT results of Matsen for cylindrical and spherical domain brushes\cite{matsen_strong-segregation_2010} in the limit of strong stretching, further supporting their validity.

Finally, in addition to depleting endpoints away from the grafting substrate, the EEZ has a strong influence in on the orientation of segments in the brush. 
To see this, consider the polar order parameter $\mathbf{p}(z)$, the average orientation of segments at point $z$ \cite{Fredrickson1992,Zhao2012,Prasad2017}.
Microscopically, the monomer orientation is tangent to the chain, i.e. $\hat{\mathbf{r}} = a^{-1}{\rm d}\mathbf{r}/{\rm d}n$, where the conformation of an individual chain is given by the parametric curve $\mathbf{r}(n)$.
For strongly stretched polymer brushes, the mean orientation of segments points along the normal direction $\mathbf{\hat{n}}$.
Fluctuations reduce the magnitude of the mean orientation, so the mean orientation of $\delta n$ monomers in a single chain is given by $\mathbf{\hat{n}}\delta z/(a\delta n)$, where $\delta z$ is the height interval spanned by $\delta n$ monomers.
Now consider averaging this quantity over all chains passing through a sample volume $\delta V$.
Once again, chain conformations are labeled by the location of the terminal end $z_0$ and the multiplicity of such chains is $\delta A \times g(z_0)$, where $\delta A$ is the cross-sectional area of the sample volume.
In general, such a volume contains $\delta n$ segments per chain.
However, we can choose the sample volume to enclose a single monomer on average, taking the limit $\delta n \rightarrow 1$ with $\delta V \simeq \delta A\times \delta z \rightarrow \rho^{-1}$.
In this limit, the polar order parameter is given by
\begin{equation}
    \mathbf{p}(z) = \frac{\mathbf{\hat{n}}}{\rho a}\int_z^h{\rm d}z_0\,g(z_0) = \mathbf{\hat{n}}\frac{h_{\rm fl}}{Na}\int_z^h{\rm d}z_0\,\hat{g}(z_0)\, .
\end{equation}
Thus, averaging over all chains in a region is equivalent to summing over chain the distribution $\hat{g}(z_0)$ from $z$ to $h$.
Note that the upper bound on the polar order parameter is set by the ratio of the flat brush height to the total arclength of a chain $h_{\rm fl}/(Na)$.
Consequently, polymer brushes locally have a zone of uniform, enhanced polar order in an EEZ, as shown in Fig.~\ref{fig:4}(c), which tapers off for $z > z_{\rm ex}$, ultimately depolarizing at the tips of the brush, i.e.~$p \to 0$ as $z\to h$.  
We note that polar order parameter for negatively curved brushes which approach the boundary between regions C and D, have polar order parameters that reflect differences in their local geometry, i.e.~for small $z$ they exhibit splay geometry near the grafting surface, but for $z \to h$ they have convergent (concave) geometry, like the inside of a cylindrical domain.  
Hence, such a brush (e.g.~the curve $H h \simeq 0.5$;  $K h^2 \simeq -2$ in Fig.~\ref{fig:4}(c)) shows both a constant polarization zone in a narrow EEZ near to the grafting surface, but also polarization profile similar to a inwardly curved cylinder near its convergent tips, with $p'(z) \to 0$ as $z \to h$, as shown in the dashed curved for the $H h = - 1/2$.

\section{\label{sec:conclusions} Conclusions}

We have presented a form of the self-consistent brush equations implementing end-exclusion zone corrections that is applicable to substrates of arbitrary curvature.
In the process, we identified several regimes of interest and developed numerical methods to approximate solutions to the equations within those regimes.
These numerical solutions provide rigorous predictions for the size of the end-exclusion zone, as well as the exact strong-stretching free energy, chemical potential field, chain end distribution, and polar order parameter of the brush as a function of substrate mean and Gaussian curvatures for arbitrary convex brush shapes ($H h>0$).  
These result show that the the magnitude of the EEZ zone, as well as its thermodynamic correction to the PBT, decreases both with mean curvature (strictly vanishing as $Hh \to 0_+)$ as well as increasingly negative Gaussian curvature for fixed $H h >0$.  
Moreover, we have demonstrated that the end-exclusion zone corrections to free energy predicted by the parabolic brush theory are non-analytic.
Consequently, while the polymer brush free energy can be well approximated by a Helfrich-style free energy for low curvature, end-exclusion zone effects cannot be added in as higher order polynomial terms in $H$ and $K$, counter to the usual expectations for structured fluid membranes \cite{MilnerBendingModuli1988,MilnerMacromolecules1994,Birshtein2008,Lei2015}.
We have shown that brush splay not only affects the distribution of free ends within the brush, but also leads to anomalous constant polarization of segments within the EEZ.

While we have neglected a discussion of solvated brushes, the results presented here can be adapted to their study.
Bringing a polymer brush into equilibrium with a solvent bath relaxes the space-filling constraint on the monomer distribution, altering eq.~(\ref{eq:incompressibility}) so that the local monomer density is $\rho A(z)\phi(z)$, where $\phi(z)$ is the volume fraction of monomers.
The volume fraction $\phi(z)$ is determined by the local chemical potential through the constitutive relation $\mu(z) \propto \phi(z)$ for marginal solvents (or $\propto \phi^2(z)$ for $\theta$-solvents) \cite{degennes_scaling}.
Solvated brushes are thus swollen relative to molten brushes, depressing the relative height of the EEZ $z_{\rm ex}/h$.
Moreover, the monomer volume fraction $\phi$ is minimal at the free surface of the brush since it is in osmotic equilibrium with the surrounding solvent bath.
Consequently, chain ends should be depleted from the free surface, altering $g(z)$ such that the maximum density of chain ends occurs in the bulk of the brush (see, e.g.~ref\cite{Belyi2004}).

While we have addressed the problem of general curvature for brushes of homogeneous, monodisperse flexible polymers, the lessons learned here prompt additional questions about more complex brushes and suggest routes towards engineering chain statistics.
The brushes we have studied have three operative length scales that control chain statistics: two radii of curvature and the brush height.  
Notably, the appearance of the an EEZ alters the shape of $\mu(z)$ thereby re-weights the local free energy cost (per unit volume) of segments in molten brushes a certain distance from the grafting surface.   
We note that similar effects are predicted for flat brushes composed of bidisperse chain lengths \cite{MWC1989,witten_two-component_1989}.
The behavior of polydisperse curved brushes is an open question, in particular whether the intrinsic stratification in differently chain lengths in polydisperse brushes either enhances, or instead suppresses, the geometric tendencies that drive EEZs in monodisperse convex brushes, and how in turn the combination of these two effects alters the thermodynamic sensitivity of brushes to curvature.

\begin{acknowledgments}
The authors are grateful to A.~Reddy and T.~Witten for valuable discussion and comments. This work was supported by the US Department of Energy
2673 (DOE), Office of Basic Energy Sciences, Division of Materials Sciences and Engineering, under Award DE-SC0014599.
\end{acknowledgments}

\section{Appendices}

\appendix

%

\section{Cylinder equations}\label{app:cyl}

Cylinders have $K=0$ and thus represent a singular limit of the general curvature equations.
Nevertheless, previous authors have solved the EEZ problem in the cylindrical limit and we can re-define the field $\psi(\tmu)$ to be well-defined in this limit.
For $K=0$, the dimensionless Steiner's equation is $\mathcal{A}(\tmu) = 1 + 2\hat{H}\hz(\tmu)$, which can be inverted to yield $\hz(\tmu) = [\mathcal{A}(\tmu) - 1]/(2\hat{H})$.
We choose $\psi(\tmu)/\Psi = \mathcal{A}(\tmu)$ so that $\hz(\tmu) = [\psi(\tmu)/\Psi - 1]/(2\hat{H})$.
Substituting $\psi(\tmu)$ into the EEZ constraint equations eqs.~(\ref{eq:EEZ1}) \& (\ref{eq:EEZ2}), we find a pair of coupled inhomogeneous linear Fredholm integral equations of the second kind,
\begin{subequations}
\begin{align} \label{eq:psi_int_equations_cyl}
\psi(\tmu < 1) &= \sqrt{1-\tmu} + \frac{2}{\pi}\Psi\arctan\sqrt{\frac{1-\tmu}{\tilde{P}-1}} \notag \\
&\mkern+32mu + \frac{1}{\pi}\int_1^{\tilde{P}} {\rm d}\tmu' \mathbbm{K}_<(\tmu,\tmu')\psi(\tmu')\\
\label{eq:psi_int_equations_cyl2}
\psi(\tmu > 1) &=  \frac{1}{\pi}\int_0^1 {\rm d}\tmu' \mathbbm{K}_>(\tmu,\tmu')\psi(\tmu')
\end{align}
\end{subequations}
where we choose $\Psi = (2\hat{H})^{-1}\sqrt{Q_{\rm fl}h^2/(Q h_{\rm fl}^2)}$.

\section{Weakly curved brushes}\label{app:weak}

To analyze the brush equations in the weak curvature limit, we follow Belyi\cite{Belyi2004} and introduce $\epsilon_P \equiv \sqrt{\tilde{P}-1}$ as a small parameter quantifying the distance to the flat brush limit.
Next, transform the integral equations through a coordinate change, introducing $u \in [0,1]$, where $u = \sqrt{1-\tmu}$ for $\tmu \leq 1$ and $u = \sqrt{(\tmu - 1)/(\tilde{P}-1)}$ for $1 < \tmu \leq \tilde{P}$, yielding
\begin{subequations}
\begin{align} \label{eq:psi_int_equations_weak1}
\psi_<(u) &= \alpha u + \frac{2}{\pi}\Psi\arctan\left(\frac{u}{\epsilon_P}\right) \notag \\
&\mkern+32mu + \frac{2\epsilon_P}{\pi}\int_0^1 {\rm d}u' \frac{u}{\epsilon_P^2u'^2 + u^2}\psi_>(u') \\
\label{eq:psi_int_equations_weak2}
\psi^2_>(u) &= \frac{2}{\pi}\Psi^2(1 - \alpha)\arctan\left(\epsilon_P u\right) \notag\\
&\mkern+32mu + \frac{2\epsilon_P}{\pi}\int_0^1 {\rm d}u' \frac{u}{\epsilon_P^2u^2 + u'^2}\psi^2_<(u')
\end{align}
\end{subequations}
where $\psi(\tmu(u) < 1) \mapsto \psi_<(u)$ and $\psi(\tmu(u) > 1) \mapsto \psi_>(u)$.
Note that the above system of equations no longer describes a coupling between two segments of the $\psi(\tmu)$ field evaluated on two subintervals of $\tmu \in [0,\tilde{P}]$, but instead describes a coupling between two separate fields, $\psi_<(u)$ and $\psi_>(u)$, on the same interval.
This facilitates an expansion of the fields as $\psi_<(u) = \Psi + \xi_<(u)$ and $\psi_>(u) = \Psi + \xi_>(u)$, where $\xi_<$ and $\xi_>$ are taken to be small so that eqs.~(\ref{eq:psi_int_equations_weak1}) \& (\ref{eq:psi_int_equations_weak2}) linearize to
\begin{subequations}
\begin{align} \label{eq:xi_int_equations1}
\xi_<(u) &= \alpha u   + \frac{2\epsilon_P}{\pi}\int_0^1 {\rm d}u' \frac{u}{\epsilon_P^2u'^2 + u^2}\xi_>(u') \\
\label{eq:xi_int_equations2}
\xi_>(u) &= -\frac{\alpha}{\pi}\arctan\left(\epsilon_P u\right) + \frac{2\epsilon_P}{\pi}\int_0^1 {\rm d}u' \frac{u}{\epsilon_P^2u^2 + u'^2}\xi_<(u')
\end{align}
\end{subequations}
where continuity in $\psi$ leads to the boundary condition $\xi_<(1) = \xi_>(1) = 0$.
Next, substitute eq.~(\ref{eq:xi_int_equations1}) into eq.~(\ref{eq:xi_int_equations1}) in retain terms up to linear order in $\epsilon_P$, yielding the integral equation
\begin{align}\label{eq:xi_final}
\xi_>(u) &= -\frac{2\alpha \epsilon_P}{\pi}u\left[\omega + \ln u\right] \notag\\
&\mkern+32mu  + \frac{4}{\pi^2}\int_0^1 {\rm d}u'\,\frac{u\ln(u/u')}{u^2 - u'^2}\xi_>(u') \, ,
\end{align}
where $\omega \equiv \Psi/2 + \ln\epsilon_P$.
Setting $\xi_>(u) = (2\alpha \epsilon_P/\pi)G(u)$, eq.~\ref{eq:xi_final} becomes
\begin{equation}\label{eq:G_Int}
G(u) = -u\left[\omega + \ln u\right] + \frac{4}{\pi^2}\int_0^1 {\rm d}u'\,\frac{u\ln(u/u')}{u^2 - u'^2}G(u') \, ,
\end{equation}
where the boundary condition $G(1) = 0$ fixes the constant
\begin{equation}
\omega = -\frac{4}{\pi^2}\int_0^1 {\rm d}u'\,\frac{\ln u'}{1- u'^2}G(u') \, .
\end{equation}
We find an numerical solution to eq.~(\ref{eq:G_Int}) through iteration, truncating after 100 iterations with a calculated residual change of $|{\rm Res}[G^{100}] - {\rm Res}[G^{99}]| \lesssim 10^{-13}$ with the result of $\omega \simeq 0.3863$, which has also been reported by Belyi\cite{Belyi2004}.

To establish the scaling of the EEZ thickness with mean curvature, note that in the limit of a disappearing EEZ, $Q \rightarrow Q_{\rm fl}$ and $h \rightarrow h_{\rm fl}$, so $\Psi \simeq \hat{H}^{-1}$.
Therefore, $\hat{H}^{-1} \simeq 2(\omega - \ln\epsilon_P)$ and
\begin{equation}
\epsilon_P \simeq e^{\omega - (2\hat{H})^{-1}} \, .
\end{equation}
Next, using the relation $\hat{z}_{\rm ex} = \left(\psi(\tmu = 1)/\Psi - 1\right)$ and the expansion $\psi(\tmu = 1) = \Psi + \xi_>(0)$, we have $z_{\rm ex}/h = (2\epsilon_P/\pi)G(0)$.
Thus, we arrive at the scaling relationship\cite{Ball1991,Belyi2004}
\begin{equation}
\frac{z_{\rm ex}}{h} \simeq A_{\rm ex}e^{-\frac{1}{2Hh}} \, ,
\end{equation}
where $A_{\rm ex} \equiv (2/\pi)G(0)\exp(\omega) \approx 0.9368$ is a fixed numerical prefactor.

To find the scaling of the EEZ-induced free energy correction $\Delta \mathcal{F}$, we start with the expression for the free energy, eq.~(\ref{eq:fe}) and use the change of variables $\mu \mapsto u$ and $\psi(\tmu) \mapsto (\xi_<,\xi_>)$, yielding
\begin{align}
F &= \frac{Q\rho A_0}{\alpha H}\int_0^1{\rm d}u\, u\bigg\{\frac{\xi_<}{\Psi}\left[1 + \frac{2}{\alpha}\frac{\xi_<}{\Psi} + \frac{1}{\alpha}\left(\frac{\xi_<}{\Psi}\right)^2\right] \notag \\
&\mkern+32mu+\epsilon_P^2\frac{\xi_>}{\Psi}\left[1 + \frac{2}{\alpha}\frac{\xi_>}{\Psi} + \frac{1}{\alpha}\left(\frac{\xi_>}{\Psi}\right)^2\right]\bigg\} \, .
\end{align} 
Substituting in eq.~(\ref{eq:xi_int_equations1}) and retaining only leading order corrections in $\epsilon_P$, we find
\begin{align}
F &\simeq \frac{Q\rho A_0}{3}\bigg\{1 + \frac{3}{2}Hh + \frac{3}{5}Kh^2 + \frac{12\epsilon_P^2}{\pi^2}\mathcal{A}(h)\int_0^1{\rm d}u\,G(u)\bigg\} \, .
\end{align}
Noting that $F_{\rm PBT}/F_{\rm fl} = \left(1+ (3/2)Hh + (3/5)Kh^2\right)$, we find 
\begin{equation}
\Delta\mathcal{F} \simeq B_{\rm ex}\mathcal{A}(h)e^{-\frac{1}{Hh}} \, ,
\end{equation}
where $B_{\rm ex} \equiv (12/\pi^2)\exp(2\omega)\int_0^1{\rm d}u\,G(u) \simeq 1.62417$ is a numerical prefactor. 
The apparent scaling $\Delta \mathcal{F} \sim (z_{\rm ex}/h)^2$ suggests that the free energy correction may be interpreted in terms of the additional chain stretching that occurs in the EEZ.

\section{The stretching free energy}\label{app:fe}

This derivation is a generalization of one introduced by Matsen \cite{Matsen2002}.
The stretching free energy of a single chain is given by
\begin{equation}
F_1(z_0) = \sqrt{\frac{3}{2a^2}}\int_0^N {\rm d}n \, \left|\frac{{\rm d}z}{{\rm d}n}\right|^2 \, ,
\end{equation} 
where we note that the conformation of the chain depends only on the location $z_0$ of its free end.
Integrating by parts, the single chain stretching free energy simplifies to
\begin{equation}
F_1(z_0) = -\sqrt{\frac{3}{2a^2}}\int_0^N {\rm d}n \, z(n)\frac{{\rm d}^2z}{{\rm d}n^2} = -\frac{1}{2}\int_0^N {\rm d}n \, z(n)\frac{{\rm d}\mu}{{\rm d}z} \, ,
\end{equation}
where we have used the equation of motion $\sqrt{3/(2a^2)}\partial_{nn}z = {\rm d}\mu/{\rm d} z$ and the boundary conditions $z(N) = 0$ and $\partial_n z|_{n=0} = 0$.
Note that this is equivalent to the virial theorem of classical mechanics.
After making a change of variables from $n$ to $z$, $F_1$ is given by
\begin{equation}
F_1(z_0) = \frac{1}{2}\int_0^{z_0} {\rm d}z \, \frac{{\rm d}n}{{\rm d}z} z\frac{{\rm d}\mu}{{\rm d}z} \, ,
\end{equation}
The free energy $F$ for the brush is given by the weighted sum of $F_1(z_0)$ over all free chain end positions $z_0$, i.e.
\begin{align}
F &= \int_0^h {\rm d}z_0 g(z_0) F_1(z_0) \notag \\
&= \frac{1}{2}\int_0^h {\rm d}z_0 \int_0^{z_0} {\rm d}z\, g(z_0) \frac{{\rm d}n}{{\rm d}z}(z|z_0) z\frac{{\rm d}\mu}{{\rm d}z} \, ,
\end{align}
where we use the notation $(z|z_0)$ to indicate that the ${\rm d}n/{\rm d}z$ depends on $z$ as well as the location $z_0$ of a chain end.
Exchanging the order of integration, the stretching free energy becomes
\begin{equation}
F = \frac{1}{2}\int_0^h {\rm d}z \left\{\int_z^h {\rm d}z_0\, g(z_0) \frac{{\rm d}n}{{\rm d}z}(z|z_0)\right\} z\frac{{\rm d}\mu}{{\rm d}z} \, ,
\end{equation}
which simplifies upon application of eq.~(\ref{eq:incompressibility}), leaving eq.~(\ref{eq:fe}),
\begin{equation}
F = -\frac{\rho}{2}\int_0^h {\rm d}z\, A(z) z \frac{{\rm d}\mu}{{\rm d}z} = \frac{\rho}{2}\int_0^P {\rm d}\mu\, z(\mu) A(\mu) \, .
\end{equation}

\bibliography{refs}

\end{document}